\documentclass{article}
\usepackage{natbib}
\usepackage{epsf}
\usepackage{epsfig}
\usepackage{amsfonts}
\usepackage{amssymb}
\usepackage{amsmath}
\usepackage{amsthm}
\usepackage{latexsym}
\usepackage{anysize}
\usepackage{graphicx}
\usepackage{subfigure}
\usepackage{url}
\usepackage{rotating}
\usepackage{multirow}
\usepackage{color}

\newcommand{\new}{\newcommand}

\def\rone{{\mathbb R}}

\def\Col{p}

\def\iter{l}

\def\M{X}
\def\MT{X^T}
\def\y{Y}
\def\b{\beta}

\def\pone{\tau}
\def\ptwo{\mu}
\def\Loss{\Lambda}

\new{\nor}[1]{\left\|{#1}\right\|}
\new{\noru}[1]{\left\|{#1}\right\|_1}

\date{}

\title{A Regularized Method for Selecting Nested Groups of Relevant Genes from
Microarray Data}
\author{Christine De Mol\\
Universit\'e Libre de Bruxelles, Dept Math. and ECARES\\
Campus Plaine CP217, 1050 Brussels, Belgium \\
telephone: + 32 2 6505573~~ fax: + 32 2 6505867~~
demol@ulb.ac.be\\
\and
Sofia Mosci\\
Universit\`a di Genova, DISI \& DIFI\\
Via Dodecaneso 35, Genova, Italy\\
telephone: +39 010 3536610~~ fax: +39 010 3536699~~
mosci@disi.unige.it \and
Magali Traskine\\
Universit\'e Libre de Bruxelles, Dept Math.\\
Campus Plaine CP217, 1050 Brussels, Belgium.\\
telephone: + 32 2 6505573~~ fax: + 32 2 6505867~~
mtraskin@ulb.ac.be\\
\and
Alessandro Verri\\
Universit\`a di Genova, DISI\\
Via Dodecaneso 35, Genova, Italy\\
telephone: +39 010 3536601~~fax: +39 010 3536699~~
verri@disi.unige.it }

\begin{document}
 
\maketitle

\begin{center}
{\bf \Large Abstract}\\ 
\end{center}
\normalsize Gene expression analysis aims at identifying the genes
able to accurately predict biological parameters like, for
example, disease subtyping or progression. While accurate
prediction can be achieved by means of many different techniques,
gene identification, due to gene correlation and the limited
number of available samples, is a much more elusive problem. Small
changes in the expression values often produce different gene
lists, and solutions which are both sparse and stable are
difficult to obtain.
We propose a two-stage regularization method able to learn linear
models characterized by a high prediction performance. By varying
a suitable parameter these linear models allow to trade sparsity
for the inclusion of correlated genes and to produce gene lists
which are almost perfectly nested. Experimental results on
synthetic and microarray data confirm the interesting properties
of the proposed method and its potential as a starting point for
further biological investigations.\\
\noindent Matlab code is available upon request.

\section{Introduction}

The extraction of relevant biological information from gene
expression data -- like disease subtype, survival time, or
assessment of the gravity of an illness -- requires the
identification of the list of the genes potentially involved in
the process, genes which need to be further scrutinized by
cross-checking the available knowledge or through additional
quantification methods. In a typical study, the size of the data
set is less than a hundred, while the dimensionality of the data
may be tens of thousands. As a consequence, feature selection,
i.e., the identification of the gene signature actually involved
in the process under study, is a formidable task for which
classical statistics (designed to deal with large sets of data
living in low-dimensional spaces) may not be well suited.\\

A great amount of supervised learning techniques have been
proposed to address the problem of feature selection -- see e.g.
the recent surveys by \citet{Dud02}, \citet{Guy03},
\citet{Saeys07} and \citet{Ver07}, and the references therein.
One usually classifies the methods into three categories: filters,
wrappers, and embedded methods. {\it Filters} implement feature
selection through a preprocessing step disconnected from the
learning phase. Examples of filters are ranking criteria where
standard statistical tests, correlation, and mutual information
are used to score each feature \citep{golub,Wes01,For03}. Filters
are also used as a preprocessing step to reduce the huge
dimensionality of feature space. The drawback of these approaches
is that the selection of features is performed beforehand and
independently of the specific required prediction or
classification task. Moreover, the selection is often made on a
univariate basis, i.e. neglecting the possible correlations
between the features.\\

On the other hand, in {\it wrappers}, the relevance of a feature
subset is determined according to prediction performance of a
given learning machine (\citep{rfe-guyon, Koh97,furlanello2003}
and references therein). However, the exploration of all subsets
of a high-dimensional feature space is in general a very demanding
task from the computational point of view.\\

Differently from wrappers and filters, where variable selection
and training are two separate processes, {\it embedded methods}
present the advantage of incorporating feature selection within
the construction of the classifier or regression model. Besides
decision trees \citep{Bre84} and boosting methods such as the
popular Adaboost \citep{Freund97}, an appealing new trend has
emerged recently namely the use of penalized methods in genomics
or proteomics. These methods consist in the minimization of a
well-defined objective function to which a penalty term is added
in order to avoid ``overfitting'', i.e. to provide some form of
``regularization'' -- or equivalently an implicit reduction of the
dimensionality of the feature space. A variety of such methods
have been proposed in the recent literature and differ by the
choice of the objective function and of the penalty term; for a
recent overview, see e.g. \citep{Seg03},\citep{Ma08} and the
references therein.\\

Particularly interesting are penalties which allow to enforce
sparsity of the model, namely to perform automatic feature
selection by assigning truly zero weights to all but a small
number of selected features. The most famous example are the
$\ell^1$-type penalties used in the so-called LASSO regression, a
name coined by \citet{Tib96} as an acronym for ``Least Absolute
Shrinkage and Selection Operator''. The use of LASSO for genomics
is also advocated e.g. in the recent papers by \citet{Ghosh05} and
\citet{Seg06}. However, a drawback of LASSO regression in the
presence of groups of correlated features is that the method is
not able to identify all members of the group. Under the name
``elastic net'' \cite{Zou05} have proposed a modification of the
LASSO method able to overcome such drawback and to identify groups
of correlated genes.\\

Building on this elastic-net regularization strategy, we propose a
two-stage method which produces gene signatures able to
effectively address prediction problems from high-throughput data
like DNA microarray. In the first stage, the method learns from
the available data a minimal set of genes the expressions of which
are best suited to accurately predict the biological parameter
related to the problem at hand. By selecting the model through the
combination of two optimization schemes, elastic net and
regularized least squares, our method leads to a model which,
unlike the elastic net alone, is characterized by both sparsity
and low bias. In the second stage, by varying a suitable
parameter, the method is able to produce models of increasing
cardinality by gradually including genes correlated with the set
of genes identified in the first stage.\\

Being formulated as a convex optimization problem, our learning
method has a sound mathematical foundation and, moreover, as we
will see, the models can be efficiently computed through simple
and easy-to-implement algorithms.  The method relies on a truly
multivariate analysis and, in contrast to the usual gene-by-gene
analyses, does not only rank genes on the basis of their
differential expression on the samples. The embedded feature
selection mechanism takes into account the correlation patterns
arising from the organization of genes in cooperating networks.\\

We show that the method gives stable results even in the presence
of data set of low cardinality. The two main features of the
proposed method are that it provides nested list of genes and that
the genes additionally included in the longer lists are correlated
with the genes of the shorter lists. Both these properties can be
very helpful when analyzing high-throughput data and might shed
light on the biological mechanisms under study.  As shown by
\citet{DDR07}, instead, the obtained models are asymptotically
equivalent in terms of prediction accuracy. The choice of which
list is the most appropriate is left to the molecular biologist
and ultimately depends on the underlying question and the
available prior knowledge.\\

The paper is organized as follows. In Section \ref{sect:meth} we
describe the method we propose for extracting nested lists of
relevant genes. In Section \ref{sect:algo} we analyze the
algorithms we developed to solve the main underlying optimization
problem and the model selection problem. Experimental results are
presented and discussed in Section \ref{sect:exp}.\\

\section{Our approach}\label{sect:meth}

In this section we first set the notation and review some basic
concepts of learning theory which are relevant to this research.
Then, we present our method and motivate our strategy for model
selection.

\subsection{Formulation of the problem}

We assume we are given a set of $n$ examples as input/output
pairs. We denote the inputs with $x_i \in \mathcal{X}=\rone^\Col$,
$i=1,...,n$; in our case the components of the vector $x_i$ are
the expressions of the $\Col$ probe sets synthesized on the chip
for each patient $i$. We note that $n$ may be about 100 or 1000
times smaller than $p$. The outputs, or corresponding responses,
are denoted with $y_i \in \mathcal{Y}$ and can be either a
discrete class label in classification problems (e.g.
discriminating between disease subtypes), or a continuous real
variable in regression problems (e.g. a measurement of some
biological parameter, survival time, or assessment of the gravity
of the illness). The problem we face is to find which of the $p$
components are needed to predict the response $y$ as accurately as
possible from any given input $x$. In our case the model
cardinality is known to be much smaller than $p$, though the
complexity of gene regulatory networks makes it difficult to
determine the number of genes actually involved in the process.\\

We restrict our attention to linear functions, or equivalently to
vectors $\b \in \rone^\Col$, modelling the relation between $x$
and $y$ as $y = \b \cdot x$. For simplicity we assume that both
$x$ and $y$ have zero mean. As customary in learning theory, the
given examples pairs are assumed to be drawn i.i.d. from a fixed
but unknown probability density $p(x,y)$ with $(x,y) \in
\mathcal{X}\times\mathcal{Y}$. Therefore, if the risk of
predicting $\beta \cdot x$ instead of $y$ is measured by $(\b
\cdot x -y)^2$, the expected risk of a given model $\b$, in the
least-squares sense, is given by
\begin{equation}
\mathcal{E}[\b] = \int_{\mathcal{X} \times \mathcal{Y}} (y-\b
\cdot x)^2\; p(x,y)\; {\rm d}x \; {\rm d}y\, .\label{eq:EXPRSK}
\end{equation}
The goal is to determine a {\em sparse} model $\b^*$, i.e. a model
of cardinality much smaller than $\Col$ -- that is, a vector
$\b^*$ with only $k$ entries different from zero (with $k << p$)
-- for which the expected risk, $\mathcal{E}[\b^*]$, takes on a
small value. We recall that the components of the model vector are
called regression coefficients or weights.

\subsection{Penalized regression methods for learning}
\label{sect:penalized}

The core of the method we propose in this paper is the
minimization of the objective function recently proposed by
\citet{Zou05} and which we write as
\begin{equation}
\frac{1}{n}\nor{Y - \M\b}_2^2  + \ptwo  \nor{\b}_2^2 + \pone\nor{\b}_1
\label{both}
\end{equation}
with $\M$ the $n\times p$ matrix such that the entry $[\M]_{ij}$
is $j$-th component $x_{i,j}$ of $x_i$ and $Y$ the $n \times 1$
vector with $[Y]_i=y_i$. In order to be consistent with the
notation in (\ref{eq:EXPRSK}) we subtract to the $j-th$ component
of $x_i$ the average $\frac{1}{n}\sum_i^n x_{i,j}$ and to $y_i$
the average $\frac{1}{n}\sum_i^ny_i$. In other words, data are
recentered with respect to their center of mass. The first term in
(\ref{both}) measures the least-squares discrepancy of the model
$\b \in \rone^\Col$ on the $n$ training examples and is the {\em
empirical risk} -- i.e. the empirical counterpart of the expected
risk (\ref{eq:EXPRSK}).\\

The second and third terms in (\ref{both}) enforce uniqueness and
numerical stability of the minimizer by penalizing respectively
the square of the Euclidean, or $\ell^2$-norm, $\nor{\b}_2^2 =
\sum_{j=1}^p \b_j^2$, and the $\ell^1$-norm, $\nor{\b}_1 =
\sum_{j=1}^p \vert \b_j \vert$, of the model vector $\b$. The
nonnegative parameters $\ptwo$ and $\pone$ are the corresponding
{\em regularization parameters}. The minimizer
$\beta_{en}(\ptwo,\pone)$ of (\ref{both}), called the {\em na\"ive
elastic net} by \citet{Zou05}, trades closeness to the data with
the size of the $\ell^2$- and $\ell^1$-norm of the solution. Before
discussing the use of (\ref{both}) in our approach, let us first
summarize the main properties of the two penalized schemes
obtained by setting either $\ptwo=0$ or $\pone=0$ in (\ref{both}).\\

{\em Ridge regression} \citep{Hoe70,Has01} -- also known as
regularized least squares \citep{Eng96,Ber98} or regularization
networks \citep{PoGi} -- avoids overfitting by controlling the
size of the model vector $\beta$, measured by its $\ell^2$-norm.
Setting $\pone = 0$ in the objective function (\ref{both}), one
gets this $\ell^2$-norm penalized regression. The unique minimizer
is then a model vector with typically all entries different from
zero. The linear computational schemes arising from this framework
are easy to implement and produce numerically stable solutions
which, for optimal values of the regularization parameter, lead to
accurate predictions. However, they tend to distribute the weights
evenly among correlated features and, thus, they are ill-suited
for performing feature selection.\\

{\em LASSO regression} \citep{Tib96} avoids overfitting by
enforcing sparsity, i.e. by favoring model vectors with only a
small number of entries different from zero. This scheme is
equivalent to $\ell^1$-norm penalized regression and is obtained
by setting $\ptwo = 0$ in the objective function (\ref{both}). In
applications in which the solution is known to depend on a
relatively small number of features, LASSO appears to be quite
appropriate. The minimizer is known to be unique (except for very
special configurations of the inputs) and stable with respect to
noise in the output data $y_i$. However, small changes in the
components of the input data $x_i$ lead to a different feature
selection, typically with no appreciable change in the overall
expected risk (or accuracy in the performance) of the obtained
model. Consequently, when the inputs are affected by noise or the
number of examples is small compared to the number of features,
the selection of the components of the model vector $\b$ might be
driven by random fluctuations.\\

Empirical evidence \citep{Zou05} indicates that the na\"ive
elastic net produces stable solutions, exhibits an interesting
grouping effect by selecting correlated features (due to the
presence of the $\ell^2$-norm term), but suffers from a quite
severe solution bias (due to the shrinkage phenomenon induced by
the $\ell^1$-norm term). Moreover, good generalization
performances are reported only for large values of the $\mu$
parameter, case in which the obtained solution is very similar to
ridge regression. In order to contrast bias and enhance the
ability of $\ell^1$-norm of promoting sparse solutions,
\citet{Zou05} proposed to rescale the coefficients and introduced
what they called the {\em elastic net}. We rely on the na\"ive
elastic net for our method but, in order to overcome its
limitations, we explore a different direction and propose an
alternative remedy.

\subsection{A two-stage method}

Our method consists of two stages. In stage I we obtain a model
with minimal cardinality and small bias by selecting the model
through the coupling of two optimization procedures. In the first
optimization procedure we perform gene selection by minimizing
(\ref{both}) for a small value of the $\ell^2$-norm parameter
$\ptwo$. The second optimization is a regularized least squares
and consists in minimizing
\begin{equation}
\frac{1}{n}\nor{Y - \tilde{\M}\tilde{\b}}_2^2  + \lambda
\nor{\tilde{\b}}_2^2, \label{ridge}
\end{equation}
in which $\tilde{\b}$ and $\tilde{\M}$ represent respectively the
weights vector $\b$ and the input matrix $\M$ restricted to the
genes selected by the first procedure. The cross-validation
protocol which we employ for model selection yields relatively
large values of $\pone$ and very small values of $\lambda$.
Consequently, for the optimal parameter pair, the solution
obtained from the first optimization selects a small number of
genes (typically characterized by a severe bias) while the
regularized least-squares minimization (\ref{ridge}) restricted to
the selected genes returns a model capable of more accurate
predictions.\\

In Stage II we aim at gradually increasing the model cardinality
by including genes correlated to the minimal set identified in
stage I. This result is achieved by running the two optimization
procedures of the previous stage for increasing values of $\ptwo$,
while keeping $\lambda$ and $\pone$ fixed to the optimal values
obtained through the cross-validation protocol mentioned above.
The resulting one-parameter family of solutions, $\{\b_{\ptwo}|
\ptwo \ge 0\}$, yields lists of relevant genes of increasing
cardinality. The experimental results reported in Subsection
\ref{modsel} show that the obtained lists, for increasing values
of $\ptwo$, are almost perfectly nested.\\

Our findings are in line with a recent work \citep{DDR07} in which
the minimization of (\ref{both}) is shown to yield consistent
estimators of the linear model for $n\to \infty$. This means that
we can find suitable sequences of parameters $\pone_n$ and
$\ptwo_n=\epsilon\pone_n$ tending to zero as $n\to \infty$, such
that we have, in probability,
$$
\mathcal{E}[\b(\ptwo_n,\pone_n)] \to ~{\rm inf}~\, \mathcal{E}
\qquad\hbox{for}\qquad
 n\to \infty.
$$
Notice that consistency, which is obtained for any value of the
parameter $\epsilon$ controlling the degree of correlation,
implies Bayes consistency, namely that the misclassification error
of $\b(\ptwo_n,\pone_n)$, $R[\b]$, converges to the Bayes risk
$R^*$, since (\citet{Bartlett:2003})
$$
(R[\b]-R^*) \le\sqrt{\mathcal{E}[\b]- {\rm inf} \mathcal{E}}.
$$

\subsection{The need for a second optimization procedure}

\citet{leng06} have shown that if the prediction accuracy is used
as a criterion to choose the tuning parameter, $\ell^1$-norm
penalized regression methods (like LASSO) provide consistent
estimates in terms of prediction accuracy but not necessarily in
terms of variable selection. With the following toy example we
provide empirical evidence of the fact that very good prediction
accuracy and correct variable selection can be both achieved by
coupling the $\ell^1$-norm penalized regression method with a
second optimization step restricted to the selected variables.\\

We consider a linear regression model $y = x\cdot \b^*+\epsilon$
in which the samples $x$ and the model $\b^*$ belong to
$\rone^{1000}$, $y \in \rone$, and $\epsilon$ represents the
noise. The training and the validation sets are built by randomly
drawing 50 and 1000 samples, respectively, from a uniform
distribution between $[-1,1]$ for each of the 1000 components. The
true model $\b^*$ has only the first three components different
from 0 while the noise is sampled from a zero-mean Gaussian
distribution with standard deviation $\sigma=0.5$.\\

\begin{figure}[h]
\includegraphics{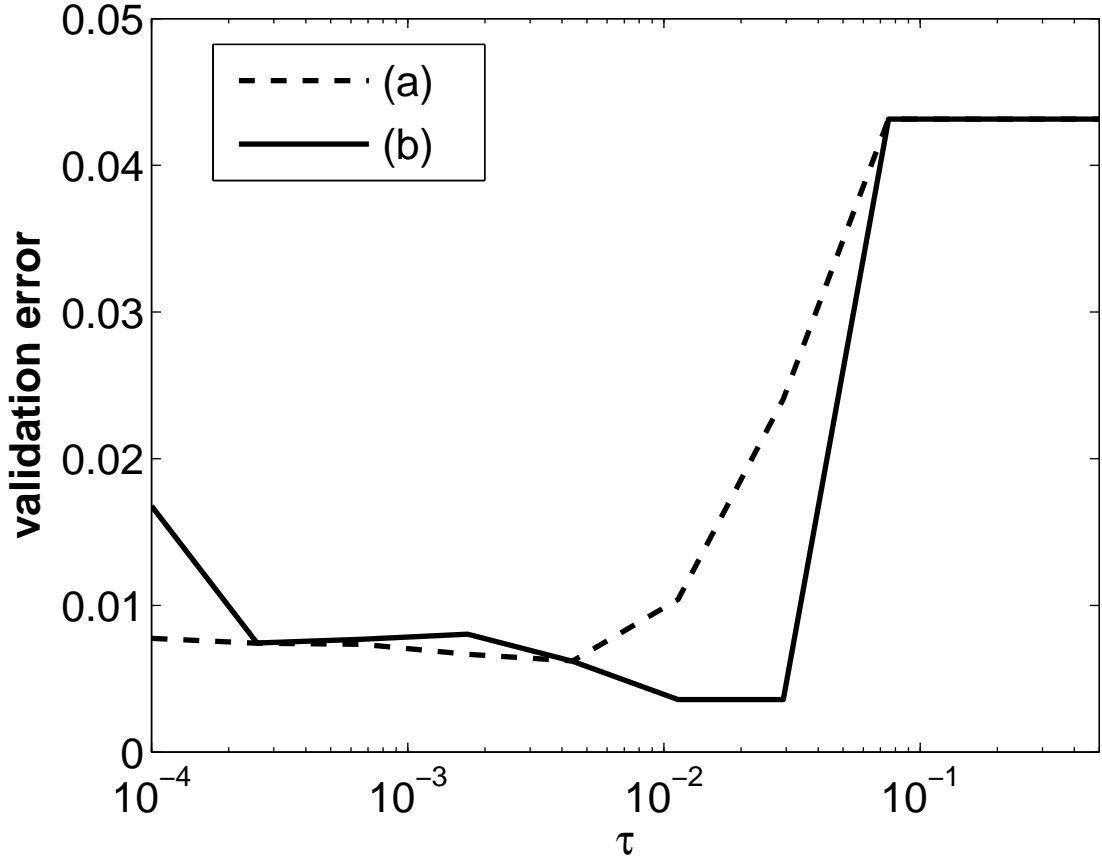}\\
\caption{\label{fig:toy}Validation error for (a) $\ell^1$-norm penalized regression, and (b) $\ell^1$-norm penalized regression
followed by ordinary least squares for different values of the penalized regression
parameter $\tau$.}

\end{figure}

For simplicity, we set $\ptwo = \lambda = 0$ and compare the results
obtained with
\begin{itemize}
\item{(a)} $\ell^1$-norm penalized regression (LASSO) alone, and
\item{(b)} $\ell^1$-norm penalized regression followed by ordinary least
squares on the selected features.
\end{itemize}
The error curves in Figure \ref{fig:toy} show how procedure (b)
allows to reach a lower minimum than (a). The validation error is
taken to be the mean-square error. The minimum is clearly reached
for a larger value of $\pone$, i.e. when the $\ell^1$-norm
penalized regression algorithm selects a lower number of features.

Let us now compare the estimators $\b^a$ and $\b^b$ obtained,
respectively, through (a) with $\pone = \pone^a$ and through (b)
with $\pone = \pone^b$ (these parameters minimize the
corresponding validation error). In Table~\ref{tab:betas} we
report the first three components of $\b^*,\b^a, \b^b$. The last
column $\b^c$, represents the output of $\ell^1$-norm penalized
regression with $\pone = \pone^b$. From Table~\ref{tab:betas}, we
can see that both $\b^a$ and $\b^b$ approximate well the relevant
components of $\b^*$. However, while $\b^b$ correctly selects the
model, $\b^a$ has many non-zero components besides the first
three. This is due to the $\ell^1$-norm penalized regression
which, in order to reduce bias, induces an optimal choice of
$\pone$ smaller than the one needed to correctly identify the
model. This can also be seen from the fact that, whereas the
$\ell^1$-norm penalized regression followed by ordinary least
squares selects the correct features (the model $\b^b$ has only
the first three components different from zero) and returns almost
perfect feature weights, the estimator $\b^c$ -- obtained for
$\pone = \pone^b$ -- underestimates all three coefficients. We can
thus conclude that the model selection obtained by coupling the
two optimization procedures appears to be more effective.

\begin{table}
\begin{center}
RELEVANT COMPONENTS OF A LINEAR MODEL IN  $\rone^{1000}$ IN A TOY PROBLEM
ESTIMATED BY DIFFERENT REGRESSION TECHNIQUES.\\
\vskip 0.3 cm
\begin{tabular}[h]{ccccccc}
 \hline
$\b^*$&&$\beta^a$&&$\beta^b$&&$\b^c$\\
 \hline
0.6449  && 0.5667  && 0.6705  && 0.3912 \\
0.8180  && 0.7389  && 0.8106  && 0.6388 \\
0.6602  && 0.5785  && 0.6794  && 0.4210 \\
 \hline
\end{tabular}
\caption{\label{tab:betas}From left to right: the true weights
of the only three non-zero components of the model
($\b^*$), the corresponding weights obtained with
$\ell^1$-norm penalized regression with $\tau=\tau^a$ ($\b^a$), with $\ell^1$-norm
penalized regression followed by ordinary least squares with $\tau=\tau^b$ ($\b^b$), and
with $\ell^1$-norm penalized regression with $\tau = \tau^b$ ($\b^c$).}
\end{center}
\end{table}


\section{Algorithmic aspects}\label{sect:algo}

In this section we discuss several algorithmic aspects of our
work. We first present an iterative algorithm for estimating the
elastic net solution, and then describe the details of our
procedure for model selection which, even for data sets of small
size, requires a very large number of optimization cycles.
Finally, we provide empirical evidence of the correctness of a
procedure capable of obtaining almost exactly the same lists of
genes with a large reduction of computing time.

\subsection{Damped iterative thresholding}

In order to minimize (\ref{both}), we use an algorithm which
generalizes the following Landweber or gradient-descent
 iterative procedure \citep{Eng96}, known to converge to a minimizer of the
unpenalized least-squares objective function $\Loss (\b)=
\frac{1}{n} \nor{\y-\M\b}_2^2$:
\begin{equation}
\b^{(\iter+1)} = \b^{(\iter)} + \frac{1}{C}[\MT \y - \MT \M
\b^{(\iter)}] ;\quad \iter = 0, 1, \dots \label{Land}.
\end{equation}
where $\MT$ denotes the transpose of $\M$ and the constant $2C$ is
a strict upper bound for the spectral norm of the matrix $\MT \M:
\Vert \MT \M \Vert < 2C$.\\

Inspired by the iterative thresholding algorithm proposed by
\citet{Dau04} for pure $\ell^1$-norm penalized regression, we propose a
double modification of Landweber algorithm which provably
converges to the minimizer of (\ref{both}). The first modification
amounts to applying a {\it soft-thresholding} operator  ${\bf
S}_{n\pone/C}$ at each iteration. The soft-thresholding operator
${\bf S}_{\alpha}$ acts on a vector component-wise as follows
\begin{equation}
[{\bf S}_{\alpha}(\b)]_j = \left\{
\begin{array}{ccl}\!\! (|\b_j|\! -\!\alpha/2)\; {\rm sign}(\b_j) &
\mbox{if} & |\b_j| \geq  \alpha/2  \\
0 & \mbox{if} & |\b_j| < \alpha/2.
\end{array} \right. \label{softthr}
\end{equation}
This operation enforces the sparsity of the regression
coefficients in the sense that all coefficients below the
threshold $\alpha/2$ are set to zero. The second modification is a
simple multiplication which leads to the following damped
iterative thresholding scheme:
\begin{equation}
\b_{en}^{(\iter+1)}\! =\! \frac{1}{1\!\!+\!\!\frac{n\ptwo}{C}} {\bf
S}_{\frac{n\pone}{C}} (\b_{en}^{(\iter)}\!+\frac{1}{C} [\MT \y \!-\!
\MT\M \b_{en}^{(\iter)}]). \label{dampedthreshLand}
\end{equation}
We recover the cases of ridge regression and damped Landweber
iteration for $\pone=0$, whereas pure $\ell^1$-regularization and
the iterative thresholding scheme considered in \citet{Dau04}
correspond to the special case $\ptwo=0$. For $\pone=\ptwo=0$, we
get the original Landweber iteration (\ref{Land}). The convergence
of (\ref{dampedthreshLand}) -- for $\ptwo>0$ and any initial
vector $\b_{en}^{(0)}$ -- to the minimizer of (\ref{both}) is a
straightforward consequence of Banach's fixed point theorem for
contractive mappings (see \citet{DDR07} for an extensive
discussion of the properties of this algorithm in a broader
setting).

\begin{figure}[h]
\begin{center}
\includegraphics{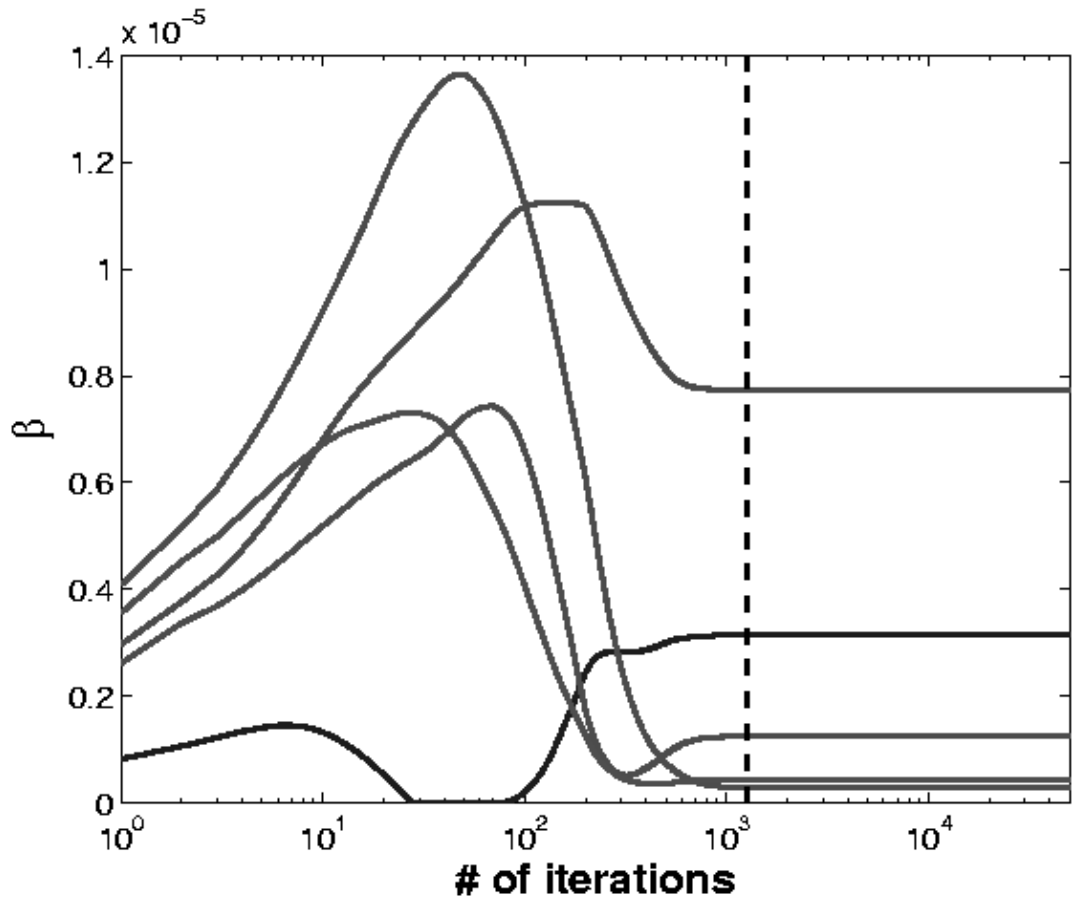}\\
\caption{\label{fig:path} Coefficient path of the relevant genes obtained
from a real data set vs. the number of iterations of
the scheme (\ref{dampedthreshLand}) on a logarithmic 
scale (the vertical dashed line corresponds to the stopping rule).}
\end{center}
\end{figure}

Since the proposed scheme is iterative we need to define a
stopping rule. We first tried to use a fixed tolerance $\delta$,
letting the iterations stop if $|\b^{(\iter+1)}_k-\b^{(\iter)}_k|_2\leq
\delta|\b^{(\iter)}_k|_2$, for all $k$. However, after extensive
experimentation on toy examples and real data, we empirically
observed that a tolerance depending on the number of iterations is
more efficient and equally easy to implement. As shown in Figure
\ref{fig:path}, if the algorithm stops when the relative change of
each coefficient $\beta^{(\iter)}_k$ is smaller than a tolerance
$\delta = 0.1/\iter$, with $\iter$ the number of performed
iterations, the support of the selected features appears to be
stabilized.

\subsection{Model selection procedure}\label{modsel}

In all of the performed experiments the training sets were
recentered as described in Subsection~\ref{sect:penalized} and the
test sample (either of the validation or of the test set) was
recentered with respect to the center of mass obtained from the
training set.\\

The data set is initially divided in
training and test set. The training set is further partitioned in
$k$ subsamples $\M_1,\dots,\M_k$ with $k$ depending on the
cardinality of the training set. In Stage I, for each subsample
$\M_i$, a classifier is first built using as training set the
remaining $k-1$ subsamples with $\pone$ and $\lambda$ ranging on a
grid in the parameters space, and then validated on $\M_i$.
Each
classifier is built by minimizing the objective functions
(\ref{both}) and (\ref{ridge}) with the current values of $\pone$
and $\lambda$ and a fixed small value for $\ptwo$ (typically
$\ptwo = 10^{-6}$). For each parameter pair the validation error
is estimated as the average error over the $k$ subsamples. Finally
the optimal parameter pair, $(\pone_{opt}, \lambda_{opt})$, is
selected as the minimizer of the validation error.\\

In Stage II a family of classifiers is built on the entire
training set with $\pone = \pone_{opt}$, $\lambda = \lambda_{opt}$
and for $m$ increasing values of $\ptwo$. Along with a test error
each classifier returns a list of variables indexed by the value
of $\ptwo$. A pseudo-code version of this procedure is summarized
in the box below.

\vskip 0.2cm
\noindent
\begin{center}
\begin{tabular}{|l|}
\hline
$~~~$\\
Given\\
-$(\M,Y)$ training set, and $(\M^{test},Y^{test})$ test set \\
-$\{(\M_1,Y_1),\dots,(\M_k,Y_k)\}$ partition of $(\M,Y)$\\
-$\ptwo_0 < \ptwo_1 < \dots < \ptwo_{m-1}$ \\
{\bf Stage I}\\
-let $\ptwo = \ptwo_0$, $(\pone_t,\lambda_l)_{t\in\mathcal{T},l\in\mathcal{L}}$ a grid in parameter space\\
-for $t\in\mathcal{T}$ and $l\in\mathcal{L}$\\
~~~~~for $i=1$ to $k$ let\\
$~~~~~~~~~X^{tr}_i:=\M_1,...\M_{i-1},\M_{i+1},...,\M_k$\\
$~~~~~~~~~Y^{tr}_i:=Y_1,...Y_{i-1},Y_{i+1},...,Y_k$\\
$~~~~~~~~~\b(t,l,i) :=$ classifier built on $(\M^{tr}_i,Y^{tr}_i)$ for
        $\pone = \pone_t,\ptwo = \ptwo_0$, and $\lambda = \lambda_l$\\
$~~~~~~~~~Err(t,l,i) :=$ error made by $\b(t,l,i)$ on $(\M_i,Y_i)$\\
~~~~~end\\
$~~~~~\overline{Err}(t,l) := \frac{1}{k}\sum_{i=1}^k Err(t,l,i)$\\
~end\\
{\bf Stage II}\\
-find $(\pone_{opt},\lambda_{opt})$ minimizing $\overline{Err}(t,l)$\\
-for $i=0$ to $m-1$ let\\
$~~~~~\b^*_{\ptwo} :=$ classifier built on $(\M,Y)$ for $\pone = \pone_{opt},
\ptwo=\ptwo_i$, and $\lambda = \lambda_{opt}$\\
$~~~~~Err^{test}_i :=$ error made by $\b^*_{\ptwo_i}$ on $(\M^{test},Y^{test})$\\
~end\\
\hline
\end{tabular}
\end{center}
\vskip 0.3cm

For small values of $\ptwo$, the solution of the damped iterative
thresholding scheme (\ref{dampedthreshLand}) -- first step for the
construction of each classifier $\beta(t,l,i)$ in Stage I --
requires a very large number of iterations. Consequently, the
procedure for determining the optimal values of $\pone$ and
$\lambda$ in Stage I, procedure which must be repeated
$k\times|\mathcal{T}|\times|\mathcal{L}|$ times with $\ptwo =
10^{-6}$, is quite time-consuming. In order to speed up the entire
process, we explored a different approach in which, for each value
of $\pone$ and $\lambda$, a series of damped iterative
thresholding schemes are run for 10 decreasing values of $\ptwo$
($\ptwo_1 = 10^{-3},...,\ptwo_{10} = 10^{-6}$), the $i$-th scheme
being restricted to the variables selected in the $(i-1)$-th
scheme with $i=2,...,10$. Extensive experiments on synthetic and
real data indicate that the features selected through this
alternative approach are almost always the same as those obtained
with the procedure described in the original scheme, but with
about a 100-fold reduction in computing time. For example
Table~\ref{table:nesting} shows the results we obtained on three
data sets of patient microarrays which we analyze in Subsection
\ref{sect:patient}. As it can easily be verified by inspection of
Table~\ref{table:nesting} the group of genes selected with the
parameter pair $(\pone_{opt},\ptwo)$ is almost perfectly enclosed
in the group selected with the parameter pair $(\pone_{opt},\bar
\ptwo)$ for $\bar\ptwo > \ptwo$. Therefore, we decided to
implement our procedure with the optimization described above.
Notice that, by doing so, the nesting of the obtained lists is
always perfect, since the optimization for $\ptwo = \ptwo_i$ is
restricted to the variables selected for $\ptwo = \ptwo_{i-1}$.

\begin{table}[!tpb]
\begin{center}
NESTING PROPERTY OF THE PROPOSED METHOD.\\
\vskip 0.3 cm
\noindent
\begin{tabular}{|lcc|lcc|lcc|}
\hline
\multicolumn{3}{|c|}{Leukemia}&
\multicolumn{3}{c|}{Lung C.}&
\multicolumn{3}{c|}{Prostate C.}\\
$  \mu          $&A  &B    &$\mu$      &A   &   B  & $\mu$      & A & B   \\
\hline
$0              $&30 &100\%&$0             $&37 &97\% &$0$      &24 &100\%\\
$6\cdot10^{-6}  $&35 &100\%&$2\cdot 10^{-7}$&36 &83\% &$10^{-6}$        &26 &100\%\\
$10^{-5}    $&38 &100\%&$3\cdot 10^{-6}$&54 &94\% &$5\cdot 10^{-6}$ &36 &100\%\\
$3\cdot10^{-5}  $&50 &100\%&$9\cdot 10^{-6}$&79 &99\% &$3\cdot 10^{-5}$ &61 & 98\%\\
$5\cdot10^{-5}  $&57 &100\%&$10^{-5}       $&98 &98\% &$4\cdot 10^{-5}$ &73 &100\%\\
$10^{-4}    $&77 &100\%&$3\cdot 10^{-5}$&152&100\%&$6\cdot 10^{-5}$ &89 & 99\%\\
$2\cdot10^{-4}  $&119&100\%&$5\cdot 10^{-5}$&182&100\%&$10^{-4}$        &123& 98\%\\
$4\cdot10^{-4}  $&144&     &$7\cdot 10^{-5}$&218&     &$4\cdot 10^{-4}$ &224&     \\
\hline
\end{tabular}
\caption{\label{table:nesting}The leftmost column contains the
values of the parameter $\ptwo$, while for each disease column A
contains the number of selected genes and B the percentage of
genes present in the gene set selected with the next
larger value of $\ptwo$ and with $\pone$ fixed at its optimal
value.}
\end{center}
\end{table}

\section{Results and discussion}
\label{sect:exp}

In this Section we report and discuss the results we obtained by
running our method on both synthetic and real data. Real-data
experiments encompass the analysis of both highly purified cell
lines grown in laboratories and samples from patients' tissues.

\subsection{A toy problem}

We first applied our method on a toy example generated according to
scenario (d) in \citet{Zou05}, where the relevant features are known
in advance. The problem is close to real gene expression data conditions
in that it encompasses both dependence on more than one variable and intra-variables correlation,
though in a setting of much lower dimensionality.\\

We consider a set of $n=100$ toy-patients. To each patient $i$ we
associate a 40-dimensional vector $x_i$ built as follows. We
divide the 40 components in four groups. Group $G_1$ consists of
the first five components $x_{i,1},...,x_{i,5}$ which are obtained
by randomly drawing a number $Z_1$ from a zero-mean Gaussian
distribution with $\sigma=1$ and adding to it a noise term
$\epsilon_j$, $j=1,...,5$, randomly drawn from a zero-mean
Gaussian distribution with $\sigma=0.01$. For each $x_{i,j}$ we
thus have
$$
x_{i,j} = Z_1 + \epsilon_j,~~ {\rm for}~ j=1,...,5.
$$
The second and third five components, belonging
to the groups $G_2$ and $G_3$ respectively, are built similarly
and read
$$
x_{i,j} = Z_2 + \epsilon_j,~~ {\rm for}~ j=6,...,10
$$
$$
x_{i,j} = Z_3 + \epsilon_j,~~ {\rm for}~ j=11,...,15.
$$
The fourth group $G_4$ consists of the remaining 25 components
randomly drawn from a zero-mean Gaussian distribution with $\sigma=1$.
By construction each of the groups $G_1$, $G_2$, and $G_3$ consists of five equivalent
variables and the true model (of all possible models the one with
largest cardinality) is written as
$$
\beta = (\underbrace{1,1,...,1}_{15},\underbrace{0,0,...,0}_{25}).
$$
For small values of $\ptwo$ the method is thus expected to select
one variable from each of the $G_1$, $G_2$, and $G_3$ groups, while for increasing values of
$\mu$ it should also include the remaining 12 variables. All the variables
of the group $G_4$, instead, should be discarded independently of $\mu$.\\

We first run stage I of the method by setting $\mu=0$ (i.e.
perform a LASSO regression) and repeat the experiment over 50 data
sets. Each data set was split in training and validation set and
the parameters $\pone^*$ and $\lambda^*$ were chosen as the ones
minimizing the error on the validation set. The method selects a
correct model (one variable from each of the three groups $G_1$,
$G_2$, and $G_3$) about 60\% of the times and a slightly redundant
set (one extra variable from any of the three groups) about 20\%
of the times.\\

\begin{figure}
\begin{center}
\includegraphics[width = \linewidth]{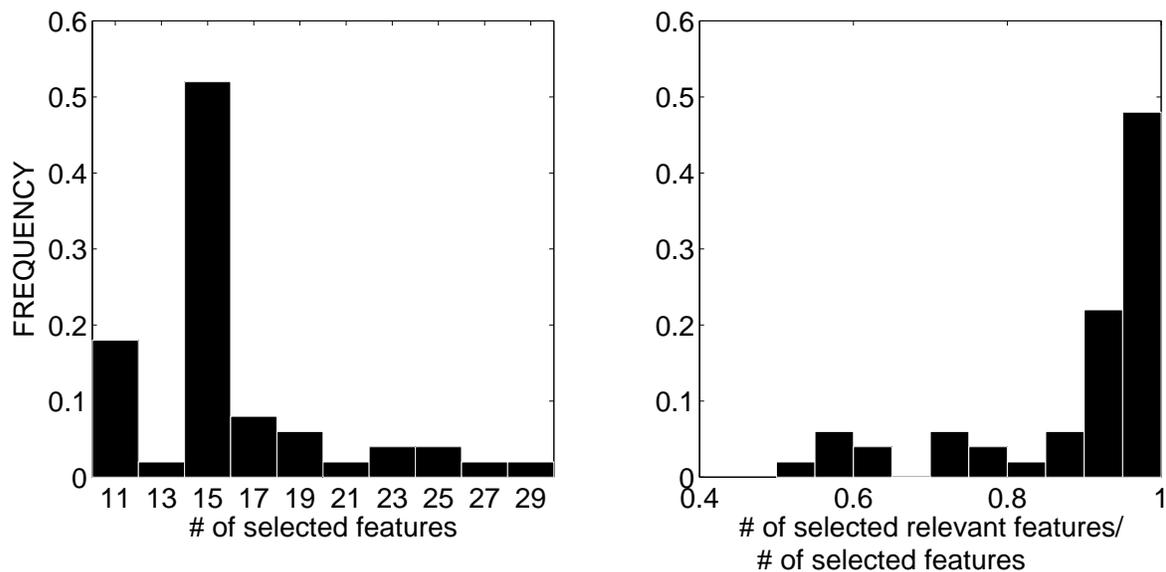}\\
\caption{\label{fig:histo_toyB}Number of variables and
accuracy of the model selected in a toy problem for $\ptwo=
1000\cdot\pone^*$; (left) frequency histogram of the number of
variables of the obtained model over 50 trials; (right) frequency
histogram of the ratio between the number of variables belonging
to either $G_1, G_2$ or $G_3$, and the overall number of selected
variables.}
\end{center}
\end{figure}

We then run stage II with $\ptwo = 1000\cdot\pone^*$.
The frequency histogram of the number of selected features is shown
in Figure \ref{fig:histo_toyB}, left. As expected, the histogram is peaked in
correspondence of 15, the maximum number of relevant variables.
The fact that the ratio between the number of the relevant features
selected by the model and the number of features selected by the model
is peaked around 1 (see Figure \ref{fig:histo_toyB}, right) confirms
that most of the times the selected features belong to the correct model.

\subsection{Cell-culture microarray data}

\begin{figure}
\begin{center}
\includegraphics[ width = \linewidth]{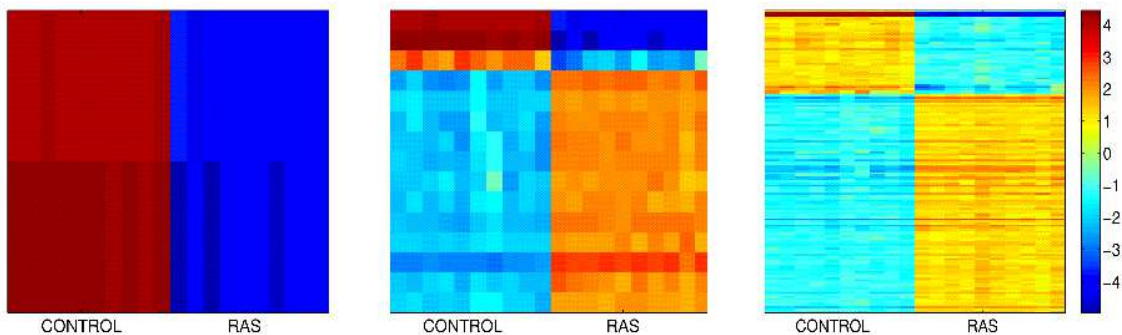}\\
\caption{\label{fig:ColorMap} Heat maps from cell cultures data.
Image intensity display of the expression levels of the 2 (left), 15 (middle)
and 144 (right) genes selected
by our method for $\ptwo = 0, ~0.05$, and $0.5$ respectively.
Expression levels are standardized to zero mean and unit variance across
samples, displayed with genes as rows and samples as columns, and colour
coded to indicate high (red) or low (blue) expression levels.}
\end{center}
\end{figure}

We now analyze the RAS data set used in \citet{bild05} and available
on line at \url{http://data.genome.duke.edu/oncogene.php}.
In \citet{bild05} human primary mammary epithelial cell cultures
(HMECs) were used to develop a series of pathway signatures
related to different oncogenes. In short, cells were infected with
different adenoviruses for eighteen hours and signatures were
extracted as the set of genes most correlated to the
classification of HMEC samples into oncogene-activated versus
control. In order to test our method for gene selection, we
applied our protocol on a subset of 20 HMEC samples, comprising
$10$ controls and $10$ samples infected with adenovirus expressing
activated H-Ras, thus extracting an alternative pathway signature
for RAS oncogene. The classification task concerned with this data
set is trivial since most classification algorithms can easily
discriminate between the two classes. In this case, however, we
are not interested in the classification performance on an
independent test set, but in the selection of relevant gene lists
and in their hierarchical structure. We thus apply our method to
the RAS data and report the heat maps of the selected gene lists
in Figure \ref{fig:ColorMap}. For $\ptwo=0$, the method extracts a
minimal set consisting of two probe sets of RAP1A (Figure
\ref{fig:ColorMap}, left), a gene belonging to the RAS oncogene
family, whereas for increasing values of $\ptwo$, the method
selects perfectly nested larger sets of genes (probe sets)
correlated  or anti-correlated with the first two, but with lower
fold change. In Figure \ref{fig:ColorMap}, middle and right, we
show the results obtained for $\ptwo=0.05$ and $\ptwo=0.5$
respectively (corresponding to $15$ and $144$ genes).\\

From the obtained results we can see that the method selects nested groups of genes
which appear to be relevant to the RAS status, nicely sorted by their differential expressions.
The two minimal probe sets are not part of the RAS signature defined in \citep{bild05},
whereas in the larger gene lists about 80\% of the genes overlap with those
found in \citep{bild05} (12 out of 15 and 112 out of 144).

\subsection{Patient-tissue microarray data}
\label{sect:patient}

Finally, we carried out experiments on data sets relative to
three diseases: leukemia, lung cancer and prostate cancer. These
gene expression data sets are available on line and concern
classification problems. The first one is the Golub data set
\citep{golub}
(\url{http://www.broad.mit.edu/cgi-bin/cancer/datasets.cgi}) which
comprises expressions of $7129$ probe sets for $72$ patients
(samples) divided in two classes according to the diagnosis of
Acute Myeloid Leukemia (AML) or Acute Lymphoblastic Leukemia
(ALL). The lung cancer data set \citep{lung}
(\url{http://www.chestsurg.org})
consists of $181$ samples with $12533$ probe sets and each
patients is either affected by malignant pleural mesothelioma
(MPM) or adenocarcinoma (ADCA). Finally, the prostate data set
\citep{prostate}
(\url{http://www-genome.wi.mit.edu/mpr/prostate})
consists again of $12533$ probe sets for $102$ samples, tumor or
normal tissue. In all cases, the vector $Y$ is formed by labels
$+1$ or $-1$ distinguishing the two classes.\\

\begin{figure}
\begin{center}
\includegraphics[width = 0.8\linewidth]{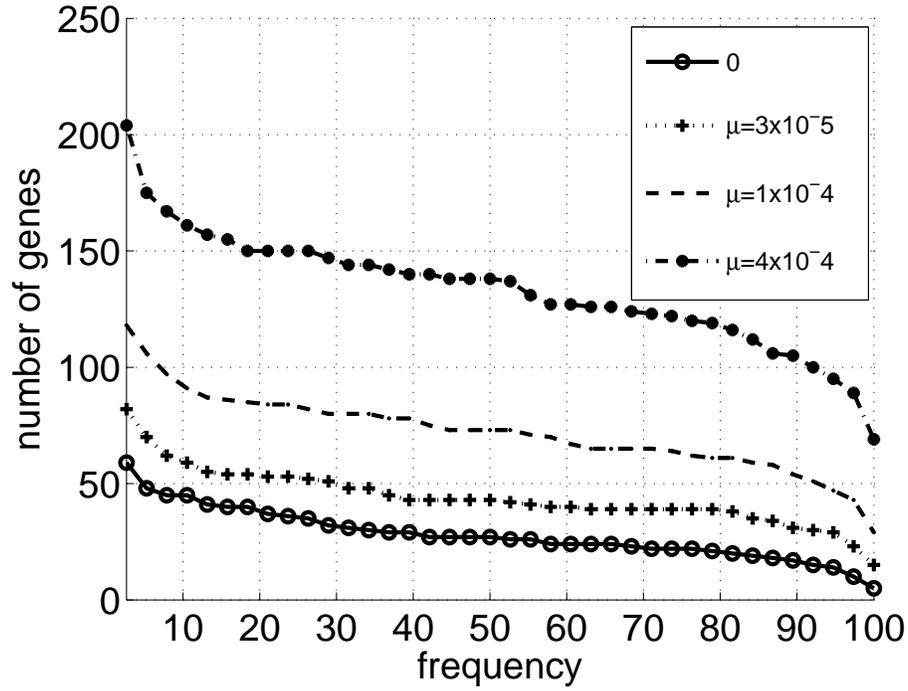}
\caption{\label{fig:freq_GOLUB} Cumulative number of
selected genes versus selection frequency in LOO cross-validation
for Leukemia data.}
\end{center}
\end{figure}

\begin{figure}
\begin{center}
\includegraphics[width = 0.8\linewidth]{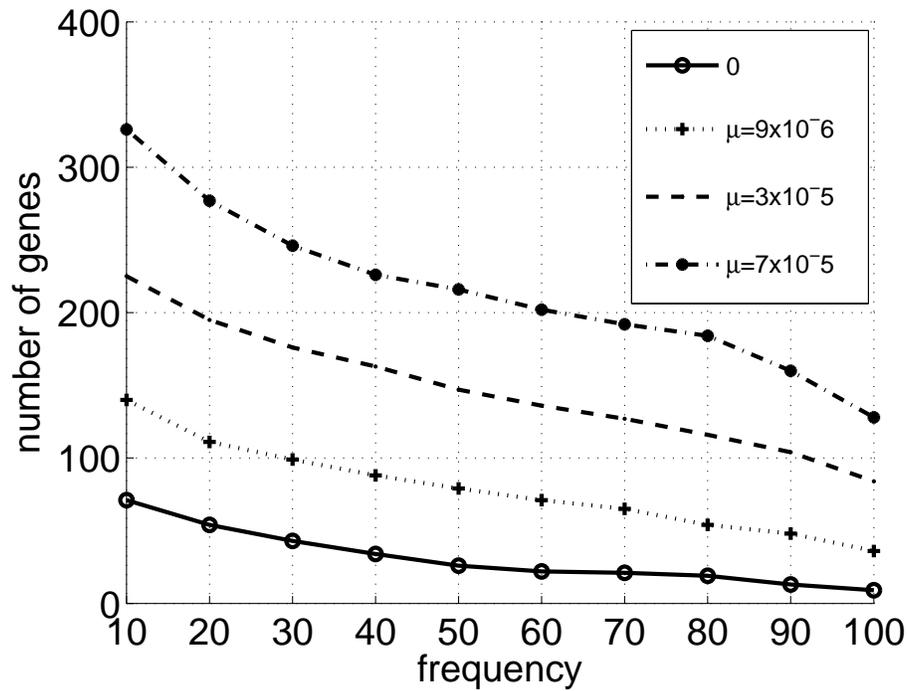}
\caption{\label{fig:freq_LUNG} Cumulative number of
selected genes versus selection frequency in $10$-fold
cross-validation for Lung Cancer data.}
\end{center}
\end{figure}

\begin{figure}
\begin{center}
\includegraphics[width = 0.8\linewidth]{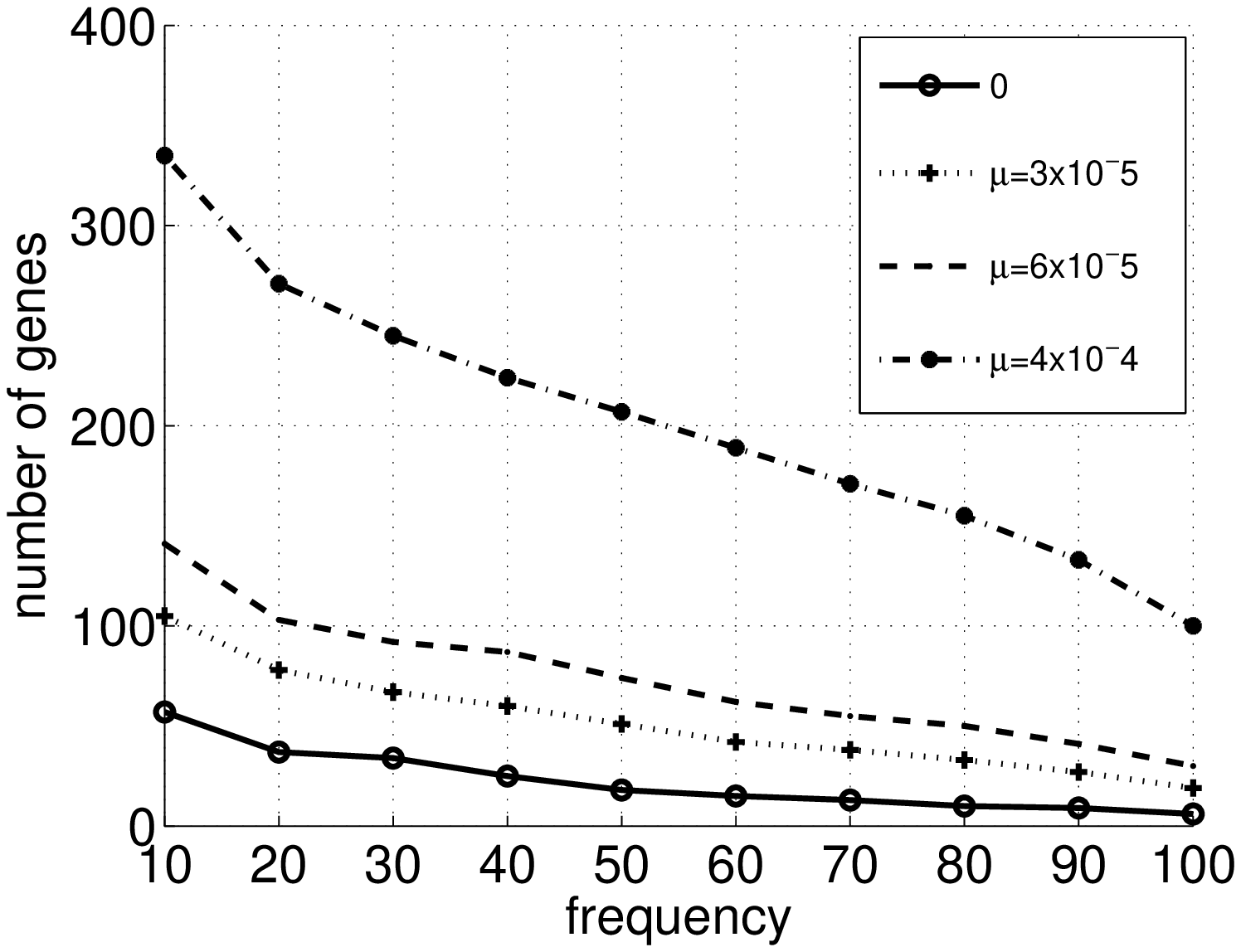}\\
\caption{\label{fig:freq_PROSTATE} Cumulative number of
selected genes versus selection frequency in $10$-fold
cross-validation for Prostate Cancer data.}
\end{center}
\end{figure}

We carried out our experiments through leave-one-out (LOO)
cross-validation on the Leukemia data (given the small number of
available samples), and 10-fold cross-validation on both the Lung
and Prostate Cancer data. A first indicator of the effectiveness
of the proposed method is the stability of the various gene lists
obtained in the training phase. In Figure \ref{fig:freq_GOLUB},
\ref{fig:freq_LUNG}, and \ref{fig:freq_PROSTATE} we report the
number of selected genes versus the selection frequency for
different values of the parameter $\ptwo$. By inspecting Figures
\ref{fig:freq_GOLUB}, \ref{fig:freq_LUNG}, and
\ref{fig:freq_PROSTATE} one sees that the produced gene lists are
remarkably stable. For increasing values of $\ptwo$ the number of
genes appearing in all of the lists ranges from about $1/3$ to
about $1/2$ of the average number of genes, while the number of
genes appearing in at least 50\% of the lists is very close to the
average.


The accuracy of the method on the three data sets (which should
remain the same for the different values of $\ptwo$) is
illustrated in Table~\ref{table:ConfLev}. By inspection we can see
that for each disease and different values of $\ptwo$ (column A)
the model cardinality from top to bottom increases (column B)
while the prediction accuracy on the test set (column B) remains
quite stable. For each disease in column B errors are reported for
the two classes separately. The rightmost column C gives the
percentage of samples which have to be rejected for both classes
in order to reach 100\% classification rate.\\

\begin{figure}
\includegraphics[height=0.48\linewidth, width = \linewidth]{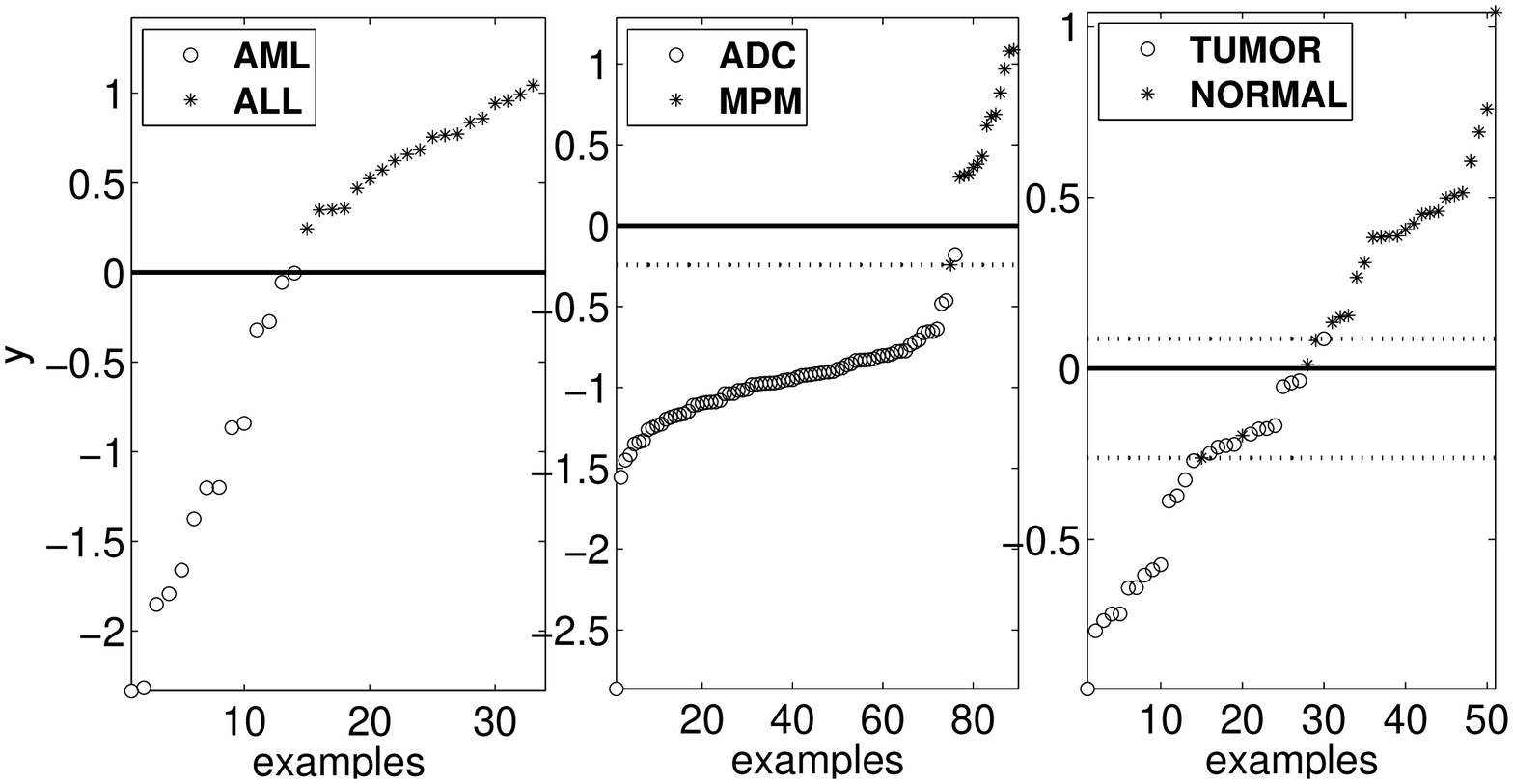}
\caption{\label{fig:CF}
Rejection region for $\ptwo=0$. In Leukemia (left) the score is perfect and
the region is degenerate, in Lung Cancer (middle) is one-sided and delimited
by the dashed line, in Prostate Cancer (right) is two sided.}
\end{figure}

The rejection region corresponding to $\ptwo = 0$ for the three
diseases is depicted in Figure~\ref{fig:CF}. The solid line gives
the decision boundary, while the dashed lines mark the rejection
region needed to reach the perfect score. No rejection region is
needed for the Leukemia study (Figure~\ref{fig:CF}, left), a
one-sided rejection region for the lung cancer study
(Figure~\ref{fig:CF}, middle) and a wider two-sided rejection
region for the prostate cancer case (Figure~\ref{fig:CF}, right).

\begin{table}[h]
\begin{center}
PREDICTION ACCURACY OF THE PROPOSED METHOD ON MICROARRAY DATA SETS.
\vskip 0.3 cm
{\footnotesize
\begin{tabular}{|rccc|rccc|rccc|}
\hline
\multicolumn{4}{|c|}{Leukemia ($n_{test} = 34$)}&
\multicolumn{4}{|c|}{Lung Cancer ($n_{test} = 90$)}&
\multicolumn{4}{|c|}{Prostate Cancer ($n_{test} = 51$)}\\
$  \mu  $   &   A   &   B   &      C        &  $ \mu $   &   A &   B &   C      &$ \mu  $        & A   & B   &      C \\
 \hline
$0$     & $28$&(0,0)&( 0\%, 0\%)&$0$         &$34$ &(1,0)&(0\%,3\%) &$0$         &$21$ &(2,1)&(13\%,48\%)\\
$6\!\cdot\!10^{-6}$&$32$ &(0,1)&( 5\%, 0\%)&$2\!\cdot\!10^{-7}$&$37$ &(1,0)&(0\%,3\%) &$10^{-6}$       &$26$ &(2,0)&(0\%,54\%)\\
$1\!\cdot\!10^{-5}$&$34$ &(0,2)&( 9\%, 0\%)&$3\!\cdot\!10^{-6}$&$51$ &(1,0)&(0\%,1\%) &$5\!\cdot\!10^{-6}$&$29$ &(2,0)&(0\%,46\%)\\
$3\!\cdot\!10^{-5}$&$40$ &(0,2)&(18\%, 0\%)&$9\!\cdot\!10^{-6}$&$78$ &(1,0)&(0\%,1\%) &$3\!\cdot\!10^{-5}$&$45$ &(3,1)&(4\%,46\%)\\
$5\!\cdot\!10^{-5}$&$50$ &(0,3)&(39\%, 0\%)&$10^{-5}$        &$108$&(1,0)&(0\%,1\%) &$4\!\cdot\!10^{-5}$&$58$ &(2,0)&(0\%,46\%)\\
$10^{-4}$   &$71$ &(1,1)&(20\%,14\%)&$3\!\cdot\!10^{-5}$&$152$&(1,0)&(0\%,1\%) &$6\!\cdot\!10^{-5}$&$72$ &(2,0)&(0\%,46\%)\\
$2\!\cdot\!10^{-4}$&$108$&(1,2)&(14\%, 8\%)&$5\!\cdot\!10^{-5}$&$174$&(1,0)&(0\%,1\%) &$10^{-4}$         &$108$&(2,0)&(0\%,50\%)\\
$4\!\cdot\!10^{-4}$&$135$&(1,2)&(14\%,15\%)&$7\!\cdot\! 10^{-5}$&$211$&(1,0)&(0\%,1\%) &$4\!\cdot\!10^{-4}$&$195$&(2,0)&(0\%,54\%)\\
\hline
\end{tabular}}
\caption{\label{table:ConfLev}For each of the three diseases the first column contains the values of the parameter $\ptwo$,
the column A the number of selected genes, the column B the number
of misclassified samples for the two original classes
respectively, and the column C the percentage of samples to be
rejected in each predicted class in order to obtain $100 \%$
classification rate. The two classes are (ALL, AML) for Leukemia,
(MPM,ADCA) for Lung Cancer, and (normal, tumor) for Prostate Cancer.}
\end{center}
\end{table}

An improvement in prediction accuracy is not the aim
of the proposed method. However, it is interesting to notice
that the proposed method reaches performances which are
at least as good as and often better than those
reported in the original studies.
In the leukemia original paper \citep{golub}, a
$50$-genes classifier is built which scored $100\%$
on the test set, though only $29$ of the $34$ test
samples corresponded to strong prediction (i.e. prediction with a
high confidence level). The prediction accuracy of our method ranges from $91\%$ to
$100\%$. As for the lung cancer data analysis in \citet{lung},
different classifiers were reported with prediction accuracy
ranging from $91\%$ to $99\%$, to be compared with the $99\%$
achieved with our algorithm. In the end, for the prostate cancer
data set, in \citet{prostate} -- after gene ranking with variation
of a signal-to-noise metric -- a $k$-NN algorithm obtained a
prediction accuracy ranging from $82.9\%$ to $95.7\%$ depending on
the number of genes used ($4$ or $6$);
with our method the accuracy ranges from $92\%$ to $96\%$.\\

Where available (leukemia and lung cancer), we have compared the
gene lists we obtained with the lists produced by other methods.
The results show partial superposition (depending on $\ptwo$) as
well as important differences.
The difference between our results and the
ones reported in the original papers is not surprising given the
multivariate flavor of our selection procedure.
Ultimately, only biological validation can assess the actual relevance
of the gene lists obtained by different methods.

\section{Conclusion}

In this paper we have proposed and analyzed a two-stage method
able to select nested groups of relevant genes from microarray
data. The first stage establishes a minimal subset of genes
relevant to the classification or regression task under
investigation. The second stage produces a one-parameter family of
groups of genes, showing a remarkable nesting property and similar
performance in terms of classification/prediction tasks. In
several problems the ability of returning nested list of relevant
genes is a key to establish the biological significance of the
obtained results and is often regarded as the most precious
information for further investigation based on biological
knowledge and subsequent experimental validation.\\

In both stages the method consists of an initial step in which a
certain amount of genes is selected through the minimization of
the objective function (\ref{both}) by means of a convergent
damped iterative thresholding algorithm and of a second step in
which the weights of the selected genes are refined through ridge
regression.\\

In the first stage, the $\ell^1$ parameter $\pone$ and the
regularization parameter $\lambda$ of the subsequent ridge
regression are estimated from the data by cross-validation, while
the $\ell^2$ parameter $\ptwo$ is set to a small value. This leads
to a solution characterized by a minimal subset of genes. In the
second stage, the $\ell^1$ and the ridge parameter are kept fixed
to their estimated optimal value and a one-parameter family of
solutions is generated for increasing values of the $\ell^2$
parameter $\ptwo$. In the proposed scheme the role of this
$\ell^2$ parameter can be thought of as a way of controlling the
trade-off between sparsity and correlation in the solution vector.\\

The results which we obtained on several data sets, including
cell-culture and patient-tissue microarray data confirm the
potential of our approach.

\section*{Acknowledgments}
We are indebted with Annalisa Barla, Ernesto De Vito, Sayan
Mukherjee and Lorenzo Rosasco for many stimulating discussions and
useful suggestions. This work has been partially supported by the
EU Integrated Project Health-e-Child IST-2004-027749, by the FIRB
project LEAP RBIN04PARL, the EU STREP grant COMBIO, the ``Action
de Recherche Concert\'ee'' Nb 02/07-281 and the VUB-GOA 62 grant.

\bibliographystyle{abbrvnat}
\bibliography{L1L2}

\end{document}